\documentclass[twocolumn,showpacs,preprintnumbers,amsmath,amssymb]{revtex4}
\usepackage{color}
\usepackage{bm, graphicx, amsmath}
\usepackage{hyperref}
\usepackage{braket}

\begin{document}

\title{Teleportation of a programmable gate}
\author{Nada Ali$^{1,2}$ and Mark Hillery$^{1,2}$}
\affiliation{$^{1}$Department of Physics, Hunter College of the City University of New York, 695 Park Avenue, New York, NY 10065 USA \\ $^{2}$Physics Program, Graduate Center of the City University of New York, 365 Fifth Avenue, New York, NY 10016}

\begin{abstract}
Alice and Bob each have a qubit, and they would like to perform a two-qubit gate on them with the assistance of Charlie.  The action of this gate depends on a parameter, which either Charlie knows, will receive later as classical information, or is encoded in a program state.  We discuss both the case in which Alice and Bob send Charlie their qubits, which he entangles with his state, and the case where Alice and Bob use teleportation to entangle their states with Charlie's state.  All of these procedures depend on the ability of Alice and Bob to apply local correction operations that depend on the results of measurements that Charlie makes, and we discuss under what conditions such operations are possible.
\end{abstract}

\maketitle

\section{Introduction}
The teleportation of quantum states makes possible the transfer of a quantum state from one quantum system to another \cite{bennett}.  As shown by Gottesman and Chuang, it is also possible to teleport quantum gates \cite{gottesman}.  As with standard teleportation, this is done using Bell-state measurements and correction operations.  Gate teleportation is employed in measurement-based quantum computation \cite{jozsa}, and has been demonstrated in several experiments.  The teleportation of CNOT and CPHASE gates has been performed optically \cite{feng,gao}.  The CNOT gate has also been teleported using trapped ions \cite{wan} and superconducting qubits \cite{chou}. Recently a Toffoli gate has been teleported using optical methods \cite{dong}.

In order to teleport a CNOT gate, one starts with two Bell pairs, $|\Phi\rangle_{12}=(1/\sqrt{2}) (|00\rangle_{12} + |11\rangle_{12})$ and $|\Phi\rangle_{34}=(1/\sqrt{2}) (|00\rangle_{34} + |11\rangle_{34})$ \cite{gottesman}.  Qubits $2$ and $3$ are sent through a CNOT gate with $3$ as the control and $2$ as the target.  Then qubits $1$and $2$ are sent to Alice, who has state $|\psi_{A}\rangle_{a} = \alpha |0\rangle_{a} + \beta |1\rangle_{a}$, and qubits $3$ and $4$ are sent to Bob, who has state $|\psi_{B}\rangle_{b} = \mu |0\rangle_{b} + \nu |1\rangle_{b}$.  Alice performs a measurement in the Bell basis on qubits $a$ and $1$, and Bob performs a measurement in the Bell basis on qubits $b$ and $4$.  They then communicate the results of their measurements to each other, and then Alice performs correction operations on qubit $2$ and Bob performs correction operations on qubit $3$.  The result is that qubits $2$ and $3$ are in the state given by a CNOT gate acting on $|\psi_{A}\rangle |\psi_{B}\rangle$, with $|\psi_{B}\rangle$ as the control and $|\psi_{A}\rangle$ as the target.  This construction will work for any gate in the Clifford group \cite{gottesman}.

Suppose we would like to implement a CPHASE($\theta$) gate, which acts as 
\begin{eqnarray}
|\psi_{A}\rangle_{a}|\psi_{B}\rangle_{b} &  \rightarrow & \alpha\mu |00\rangle_{ab}+\alpha \nu |01\rangle_{ab} \nonumber \\
&& \beta \mu |10\rangle_{ab} + e^{i\theta} \beta \nu |11\rangle .
\end{eqnarray}
The above construction will not work to teleport this gate for a general $\theta$ (it does work for $\theta = \pi$ - that is just the CPHASE gate).  We would like to find a way to teleport this gate.  Furthermore, suppose $\theta$ is unknown to the parties performing the teleportation but is suppled in the form of a program state.  How can we use the program state to accomplish the teleportation of the CPHASE($\theta$) gate?

We begin by looking at a simplified problem in which Alice and Bob send their qubits to Charlie, who then entangles them with his system and then returns the qubits to Alice and Bob. At a later time, Charlie is able to implement a CPHASE($\theta$) gate in one of two ways.  He can use a gate that applies a phase to his system and then perform measurement, or he can use a program state in which the phase is encoded, transfer the phase to his system, and then make a measurement to perform the CPHASE($\theta$) gate on Alice's and Bob's qubits.  In both cases, he tells Alice and Bob the results of his measurement, and they perform local correction operations on their qubits to remove unwanted phases.  We next examine what kinds of phase operations can be performed that allow Alice and Bob to perform local corrections and the requirements this imposes on Charlie's system.  Next we remove the requirement that Alice and Bob send their qubits to Charlie, and, instead, they make use of teleportation to send their states.  Finally, we present an implementation of one of our protocols on a quantum computer and compare the remote implementation of a CPHASE($\theta$) gate to that of a direct implementation.

\section{Simplified problem}

\subsection{Examples}

Consider the following scenario.  Alice has the state $|\psi_{A}\rangle_{a}$ and Bob has the state $|\psi_{B}\rangle_{b}$.  They send their states to a controller, whom we shall call Charlie, and he entangles them with his state, which we shall call $|\phi\rangle_{c}$, and them returns them to Alice and Bob.  At a later time Charlie modifies his state to incorporate the information about the angle $\theta$, performs a measurement that decouples his state from those of Alice and Bob, and tells both of them the result of his measurement.  Alice and Bob each perform local correction operations, and the result is that the CPHASE($\theta$) gate has been applied to the states held by Alice and Bob.  The information about $\theta$ can be possessed by Charlie at the beginning of the protocol, can be sent to him in classical form at a later time, or can be encoded in a quantum program state.

Let us now see how this works.  Charlie has a qutrit, whose basis states are $\{ |j\rangle_{c}\, | \, j=0,1,2\}$.  He combines it with the states of Alice and Bob to yield
\begin{eqnarray}
\label{eq1}
|\Psi\rangle_{cab} & = &  \alpha\mu |0\rangle_{c} |00\rangle_{ab} + |1\rangle_{c} ( \alpha\nu |01\rangle_{ab} \nonumber \\
&& + \beta\mu |10\rangle_{ab}) + \beta\nu |2\rangle_{c} |11\rangle_{ab} .
\end{eqnarray}
This state can be prepared using Controlled-SHIFT operations with the qubits as the controls and the qutrit as the target \cite{hillery}.  Charlie now returns Alice's qubit to her and Bob's qubit to him.  Charlie, at a later time, can now choose $\theta$, perhaps on the basis of classical information communicated to him, and implement the CPHASE($\theta$) gate by using a unitary operator.  Charlie applies the operator $U_{c}$ to his qutrit.  This operator acts as $U_{c}|0\rangle_{c} = |0\rangle_{c}$, $U_{c}|1\rangle_{c} = |1\rangle_{c}$, and $U_{c}|2\rangle_{c} = e^{i\theta} |2\rangle_{c}$.  Charlie then measures his qutrit in the basis
\begin{equation}
|u_{j}\rangle_{c} = \frac{1}{\sqrt{3}} \sum_{k=0}^{2} e^{2\pi ijk/3} |k\rangle_{c} ,
\end{equation}
for $j=0,1,2$.  If he obtains $|u_{j}\rangle_{c}$ the state is
\begin{eqnarray}
\alpha\mu |00\rangle_{ab} + e^{-2\pi ij/3} ( \alpha\nu |01\rangle_{ab} \nonumber \\
+ \beta\mu |10\rangle_{ab}) + \beta\nu e^{-4\pi ij/3} e^{i\theta} |11\rangle_{ab} .
\end{eqnarray}
Note that if $j=0$ the state is just the one that would be obtained by sending the original qubits of Alice and Bob through a CPHASE($\theta$) gate.  In general, if Charlie gets $|u_{j}\rangle_{a}$, then Alice and Bob can each apply the unitary operator $U= |0\rangle\langle 0| + e^{2\pi ij/3} |1\rangle\langle 1|$, which will remove the unwanted phase factors and result in an output state that is the CPHASE($\theta$) gate acting on Alice's and Bob's states.

It is also possible for the angle $\theta$ to be encoded in a program state.  We shall call the program state $|\theta_{d}\rangle$, and it is contained in a space $\mathcal{H}_{d}$, which has dimension $N$.  The state $|\theta\rangle_{d}$ is given by
\begin{equation}
\label{program}
|\theta\rangle_{d} = \frac{1}{\sqrt{N}} \sum_{j=0}^{N-1} e^{-ij\theta} |j\rangle_{d} .
\end{equation}
Send Charlie's qutrit and $|\theta\rangle_{d}$ through a Controlled-SHIFT gate that takes $|0\rangle_{c}|j\rangle_{d}\rightarrow |0\rangle_{c}|j\rangle_{d}$, $|1\rangle_{c}|j\rangle_{d}\rightarrow |1\rangle_{c}|j\rangle_{d}$, and $|2\rangle_{c}|j\rangle_{d}\rightarrow |2\rangle_{c}|j+1\rangle_{d}$, where the addition is modulo $N$.  The result is
\begin{eqnarray}
\frac{1}{\sqrt{3}}\left( \sum_{k=0}^{2} |k\rangle_{c}\right) |\theta\rangle_{d} & \rightarrow & \frac{1}{\sqrt{3N}} (|0\rangle_{c} + |1\rangle_{c} \nonumber \\
&&+ e^{-i(N-1)\theta} |2\rangle_{c} ) |0\rangle_{d} \nonumber \\
&& +  \frac{1}{\sqrt{3N}} (|0\rangle_{c} + |1\rangle_{c} + e^{i\theta} |2\rangle_{c} ) \nonumber \\
&& \left( \sum_{j=1}^{N-1} e^{-ij\theta} |j\rangle_{d}\right) .
\end{eqnarray}
Charlie can now measure the projection $I_{d}-|0\rangle_{d}\langle 0|$ on $\mathcal{H}_{d}$.  If he obtains the eigenvalue $1$, which happens with a probability of $(N-1)/N$, he obtains the state
\begin{equation}
 \frac{1}{\sqrt{3}} (|0\rangle_{c} + |1\rangle_{c} + e^{i\theta} |2\rangle_{c} ) \frac{1}{\sqrt{N-1)}} \left( \sum_{j=1}^{N-1} e^{-ij\theta} |j\rangle_{d}\right) .
 \end{equation}
 Note that if the qutrit had been entangled with Alice's and Bob's qubits as in Eq.\ (\ref{eq1}), then the resulting state in $\mathcal{H}_{abc}$ would have been the same as that obtained by applying $U_{c}$ to Eq.\ (\ref{eq1}). The program state can be reused, but with a smaller probability of success.  Note that Charlie does not need to know what $\theta$ is.  It should be mentioned that the protocol summarized here is related to one for remote state preparation for multiple receivers \cite{hillery,sukeno}.
 
\begin{figure}
\begin{center}
\includegraphics[width=24em]{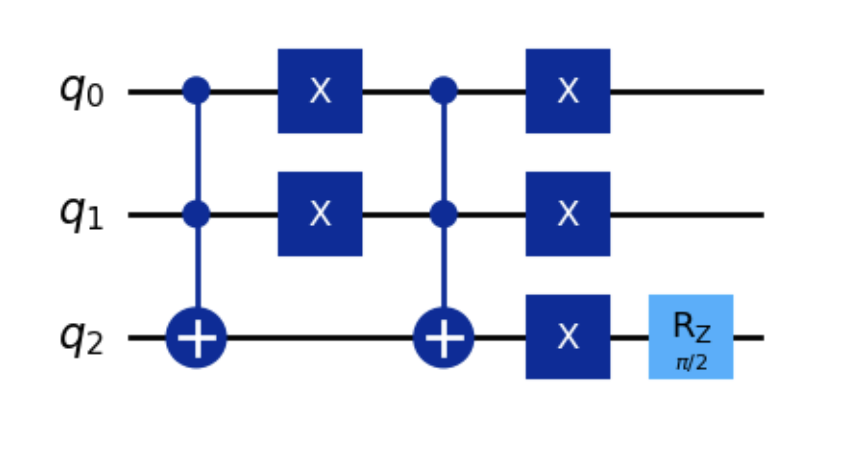}
\end{center}
\caption{Circuit for producing the state in  Eq.\ (\ref{qubit-state}).  The top line is Alice's qubit, the second line is Bob's qubit, and the third line is Charlie's qubit, initially in the state $|0\rangle$.  The state in Eq.\ (\ref{qubit-state}) appears after the three $X$ gates, and the final gate, $R_{z}$, is used to apply the phase factor $e^{i\theta}$ to Charlie's $|1\rangle$ state.}
\end{figure}

 It is also possible to apply factors of $e^{i\theta}$ to other combinations of states rather than just the $|11\rangle$ state.  For example, suppose we want to apply it to the $|01\rangle$ and $|10\rangle$ states.  In that case, Charlie only requires a qubit instead of a qutrit.  After Alice and Bob have sent Charlie their qubits and he has entangled them with his qubit, the state is
 \begin{eqnarray}
 \label{qubit-state}
 |\Psi\rangle_{cab} & = & |0\rangle_{c} (\alpha\mu |00\rangle_{ab} + \beta \nu|11\rangle_{ab}) \nonumber \\ 
&& + |1\rangle_{c} (\alpha\nu |01\rangle_{ab} + \beta \mu |10\rangle_{ab})  .
\end{eqnarray}
Charlie now returns Alice's and Bob's qubits to them.  When Charlie wants to implement the gate, he  applies the operator $R_{z}|0\rangle_{c}=|0\rangle_{c}$ and $R_{z}|1\rangle=e^{i\theta}|1\rangle_{c}$ and then, measures in the $|\pm x\rangle$ basis.  If he get $|+x\rangle$  the process is complete, and if he gets $|-x\rangle$, then Alice and Bob can each apply $\sigma_{3}$ to correct the state. 

 It is also possible in this case for the angle $\theta$ to be encoded in a program state. We will look at the case of a two-qubit program, which will allow Charlie to transfer the phase to Alice and Bob with a probability of $2/3$.  The program state is contained in a two-qubit register $(d,e)$ encoding a 3-level alphabet (see Fig. 2). The program state $\ket{\Phi}_{de}$ is given by
\begin{equation}
\label{eq:prog}
\ket{\Phi}_{de} = \frac{1}{\sqrt{3}}\left(\ket{01}_{de} + e^{-i\theta}\ket{11}_{de} + e^{-2i\theta}\ket{10}_{de}\right).
\end{equation}
We use the same entanglement circuit as in the previous example to create the Alice-Bob-Charlie state
\begin{eqnarray}
\label{eq:psi_cab}
\ket{\Psi}_{cab} & = & \ket{0}_c(\alpha \mu\ket{00}_{ab} + \beta \nu \ket{11}_{ab}) \nonumber \\
&& + \ket{1}_c(\alpha \nu \ket{01}_{ab} + \beta \mu\ket{10}_{ab}),
\end{eqnarray}
shown in Fig. 2. The key difference from before is that we do not apply the $Rz$ gate. Instead, we tensor this state with the program state and apply two Toffoli gates.  These gates implement a cyclic permutation on qubits $(d,e)$ when Charlie's control qubit $c$ is in state $\ket{1}$, and do nothing when $c$ is in state $\ket{0}$.

\begin{figure}[h]
\centering
\includegraphics[width=0.5\textwidth]{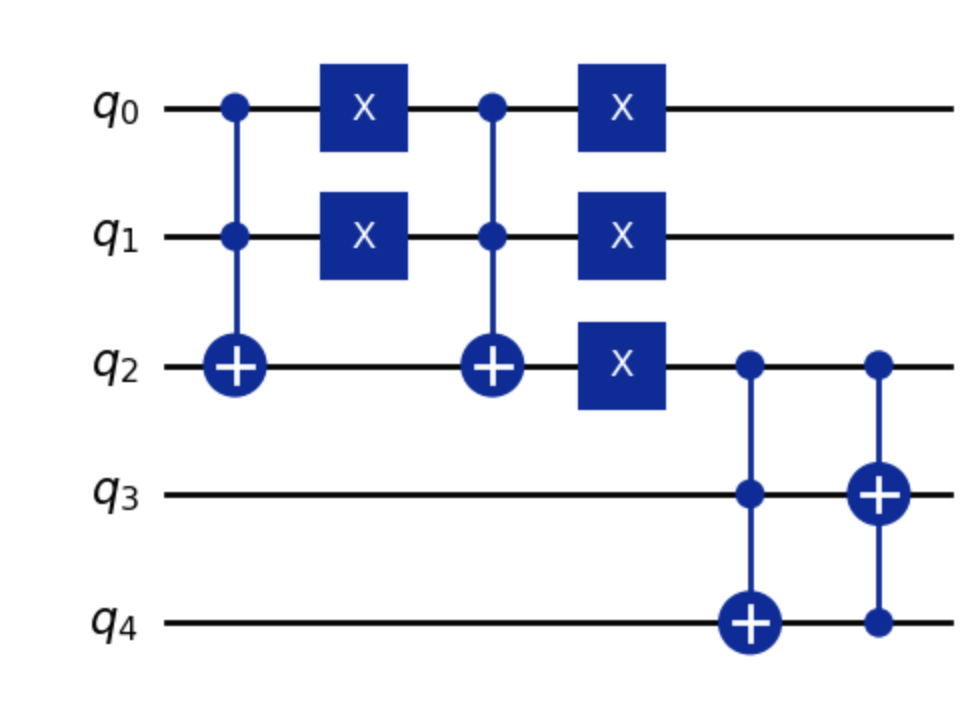}
\caption{Circuit for Example 2. Here Alice's (a) qubit is $q_{0}$, Bob,s (b) is $q_{1}$ and Charlie's (c) is $q_{2}$.  The qubits $q_{3}$ and $q_{4}$ comprise the program state $(d,e)$.  The entanglement circuit creates the state in Eq.~\eqref{eq:psi_cab}. The two Toffoli gates on the right implement a cyclic permutation on $(d,e)$ when $c=1$: the first Toffoli has controls $\{c,d\}$ and target $e$; the second has controls $\{c,e\}$ and target $d$. When $c=1$, this maps $\ket{01} \to \ket{11} \to \ket{10} \to \ket{01}$ while leaving $\ket{00}$ fixed; when $c=0$ it does nothing.}
\label{Fig2}
\end{figure}

After applying the two Toffoli gates, the combined state becomes:
\begin{eqnarray}
\ket{\Psi_{\mathrm{out}}} &= & \frac{1}{\sqrt{3}}\left[\ket{0}_c(\alpha \mu\ket{00}_{ab} + \beta \nu\ket{11}_{ab}) \right. \nonumber \\
&& \left(\ket{01}_{de} + e^{-i\theta}\ket{11}_{de} + e^{-2i\theta}\ket{10}_{de}\right) \nonumber \\
&& + \ket{1}_c(\alpha \nu\ket{01}_{ab} + \beta \mu\ket{10}_{ab}) \nonumber \\
 && \left.  \left(\ket{11}_{de} + e^{-i\theta}\ket{10}_{de} + e^{-2i\theta}\ket{01}_{de}\right)\right].
\end{eqnarray}
Note that when $c=\ket{1}$, the program state has been cyclically permuted: $\ket{01} \to \ket{11}$, $\ket{11} \to \ket{10}$, and $\ket{10} \to \ket{01}$.  Regrouping by the $(d,e)$ basis states:
\begin{eqnarray}
\ket{\Psi_{\mathrm{out}}} &= &\frac{1}{\sqrt{3}}\left( |\phi_{01}\rangle_{abc} |01\rangle_{de} + |\phi_{11}\rangle_{abc} |11\rangle_{de} \right. \nonumber \\
&& \left. + |\phi_{10}\rangle_{abc} |10\rangle_{de} \right) , 
\end{eqnarray}
where
\begin{eqnarray}
|\phi_{01}\rangle_{abc} & = & \ket{0}_c(\alpha \mu\ket{00}_{ab} + \beta \nu\ket{11}_{ab}) \nonumber \\
&& + e^{-2i\theta}\ket{1}_c(\alpha \nu\ket{01}_{ab} + \beta \mu\ket{10}_{ab}) , \nonumber \\
|\phi_{11}\rangle_{abc} & = & e^{-i\theta}\ket{0}_c(\alpha \mu\ket{00}_{ab} + \beta \nu\ket{11}_{ab}) \nonumber \\ 
&&+ \ket{1}_c(\alpha \nu\ket{01}_{ab} + \beta \mu\ket{10}_{ab}) ,
\end{eqnarray}
and 
\begin{eqnarray}
|\phi_{10}\rangle_{abc} & = & e^{-2i\theta}\ket{0}_c(\alpha \mu\ket{00}_{ab} + \beta \nu\ket{11}_{ab}) \nonumber \\
&&+ e^{-i\theta}\ket{1}_c(\alpha \nu \ket{01}_{ab} + \beta \mu\ket{10}_{ab}) .
\end{eqnarray}
Charlie now measures the program register $(d,e)$ in the computational basis. With probability $2/3$, he obtains an outcome where the phase $e^{i\theta}$ has been correctly transferred to the $\ket{1}_c$ branch. If he obtains outcome $\ket{11}_{de}$, the state becomes $|\phi_{11}\rangle_{abc}$ which, up to the global phase $e^{-i\theta}$, yields the desired state. If he obtains outcome $\ket{10}_{de}$, the state is $|\phi_{10}\rangle_{abc}$ which also yields the correct relative phase $e^{i\theta}$ on the $\ket{1}_c$ branch, up to a global phase. Therefore, these two outcomes yield the target state:
\begin{eqnarray}
\ket{0}_c(\alpha \mu\ket{00}_{ab} + \beta \nu\ket{11}_{ab}) \nonumber \\
+ e^{i\theta}\ket{1}_c(\alpha \nu\ket{01}_{ab} + \beta \mu\ket{10}_{ab}).
\end{eqnarray}

Note that this is exactly the same state as obtained by applying $R_z(\theta)$ in the previous example. The key advantage is that Charlie does not need to know the value of $\theta$: it is encoded quantum mechanically in the program state phases $1, e^{-i\theta}, e^{-2i\theta}$ and is transferred to the $\ket{1}_c$ branch through the cyclic permutation. With probability $2/3$, the protocol succeeds without any corrections.

In the case that Alice and Bob applied correction operations, note that these are local operations at Alice's and Bob's sites.  If instead we wanted to apply the $e^{i\theta}$ to the $|11\rangle_{ab}$ state with Charlie having only a qubit, then it seems we cannot do this locally.  In that case, at the end of the entangling and application of $R_{z}$ we would have the state
\begin{eqnarray}
|0\rangle_{c} (\alpha\mu |00\rangle_{ab} + \alpha\nu |01\rangle_{ab} + \beta \mu |10\rangle_{ab})) \nonumber \\ 
 +e^{i\theta} |1\rangle_{c} ( \beta \nu |11\rangle_{ab}).
\end{eqnarray}
If Charlie now measures in $|\pm x\rangle$ basis and gets $|-x\rangle$, then Alice and Bob cannot flip the sign of $|11\rangle_{ab}$ only, which is necessary to correct the state, with local measurements.  This suggests that there is a relation between the dimension of Charlie's system and the operations that can be implemented using only local correction operations.  This we will now explore.

\subsection{When local corrections are possible}

Suppose we have $n$ parties, and they want to implement an $n$-qubit gate on their qubits.  The gate will have the effect of multiplying sets of $n$-qubit states by phase factors of the form $e^{i\theta}$.  The parties all send their states to Charlie, who possess an $n$ qubit state of the form
\begin{equation}
|\psi\rangle_{a} = \sum_{k=0}^{n} |n;k\rangle_{a} , 
\end{equation}
where $|n;k\rangle_{a}$ is an unnormalized state that is a linear combination of $n$-qubit basis states, where in each basis state $k$ of the qubits are in the state $|1\rangle$.  Charlie has an auxiliary $m$-dimensional system, where $m<n$, with basis elements $\{ |r\rangle_{c}\, | \, r=0,1, \ldots, m-1\}$,  and he wants to entangle this system with the $n$ qubits.  Before looking at that, it is useful to define the $m$-dimensional basis $\{ |u_{s}\rangle \, |\, s=0,1,2, \ldots m-1 \}$, where
\begin{equation}
|u_{s}\rangle_{c} = \frac{1}{\sqrt{m}} \sum_{r=0}^{m-1} e^{2\pi irs/m} |r\rangle_{c} .
\end{equation}
Now suppose suppose that Charlie's entangling operation produces the state 
\begin{equation}
\sum_{q=0}^{m-1} \sum_{ \{ k|\,k=q\, {\rm mod}\, m\} }  |n;k\rangle_{a}|q\rangle_{c} .
\end{equation}
At this point he sends the qubits back to the parties from which they came.  He can now apply a unitary operation to the auxiliary system that introduces the phases $e^{i\theta_{q}}$, $q= 0,1,\ldots ,m-1$, resulting in the state
\begin{equation}
\sum_{q=0}^{m-1} e^{i\theta_{q}} \sum_{ \{ k|\,k=q\, {\rm mod}\, m\} }  |n;k\rangle_{a}|q\rangle_{c} .
\end{equation}
Charlie now measures the $c$ part of this state in the basis $\{ |u_{s}\rangle \}$, and obtain the result $s$.  The resulting state of the $n$ qubits is
\begin{equation}
\sum_{q=0}^{m-1} e^{i\theta_{q}} \sum_{\{ k|\, k=q\, {\rm mod}\, m\} } e^{-2\pi iqs/m}  |n;k\rangle_{a} .
\end{equation}
We now want to see if we can get rid of the phases $e^{-2\pi iqs/m}$ by local operations on the qubits, that is by having each party holding a qubit perform an operation on their qubit alone.  Let each party apply the operation $U_{s}|0\rangle = |0\rangle$ and $U_{s}|1\rangle = e^{\phi_{s}}|1\rangle$ to their qubit.  Note that these are local operations.  The angle $\phi_{s}$ depends on $s$, and Charlie broadcasts the result of his measurement to all parties.  The resulting state is
\begin{equation}
\sum_{q=0}^{m-1} e^{i\theta_{q}}  \sum_{\{ k|\, k=q\, {\rm mod}\, m\} } e^{-2\pi iqs/m} e^{ik\phi_{s}}  |n;k\rangle_{a} .
\end{equation}
Now choose $\phi_{s} = 2\pi s/m$.  We then have, for $k=q$ mod $m$, that $e^{ik\phi_{s}} = e^{2\pi i qs/m}$, and we see that the phase factors the state picked up from the measurement are cancelled.  Therefore, we can remotely apply the operation
\begin{equation}
\sum_{q=0}^{m-1} \sum_{ \{ k|\, k=q\, {\rm mod}\, m \} } |n;k\rangle_{a} \rightarrow \sum_{q=0}^{m-1} e^{i\theta_{q}} \sum_{ \{ k|\, k=q\, {\rm mod}\, m\} }  |n;k\rangle_{a} .
\end{equation}

So far the measurement bases have been inspired by representations of cyclic groups.  This can be generalized by making use of representations of products of cyclic groups.  Consider the product space $\mathcal{H}_{c1}\otimes \mathcal{H}_{c2}$, where $\mathcal{H}_{c1}$ has dimension $m$ and $\mathcal{H}_{c2}$ has dimension $n$, and the orthonormal basis for this space defined by
\begin{equation}
|X_{rs}\rangle = \frac{1}{\sqrt{mn}} \sum_{j=0}^{m-1}\sum_{k=0}^{n-1} e^{2\pi ijr/m} e^{2\pi iks/n} |j\rangle_{c1} |k\rangle_{c2} ,
\end{equation}
for $r=0, 1, \ldots m-1$ and $s=0, 1, \ldots n-1$.  Here $\{ |j\rangle_{c1}\, | \, j=0, 1, \ldots m-1\}$ is an orthonormal basis for $\mathcal{H}_{c1}$ and $\{ |k\rangle_{c2}\, | \, k=0, 1, \ldots n-1\}$ is an orthonormal basis for $\mathcal{H}_{c2}$.  

Suppose now that Charlie starts with the state
\begin{equation}
\sum_{j=0}^{m-1}\sum_{k=0}^{n-1} |j\rangle_{c1} |k\rangle_{c2} |Q_{jk}\rangle ,
\end{equation}
where $|Q_{jk}\rangle$ is a state of $m+n-2$ qubits in which the qubits are divided into two groups, one of size $m-1$ and the other of size $n-1$.  To be specific, let's suppose that the first $m-1$ qubits are in group 1 and the rest are in group 2.  The state $|Q_{jk}\rangle$ is a linear combination of states in the computational basis in which in group 1, $j$ qubits are in the state $|1\rangle$ and the rest are in the state $|0\rangle$, and in group 2, $k$ qubits are in the state $|1\rangle$ and the rest are in the state $|0\rangle$.  These qubits were sent to Charlie by $m+n-2$ external parties, Charlie has entangled them with his state and then returned the qubits to the parties.  Charlie now applies the phase factors $e^{i\theta_{jk}}$ resulting in the state
\begin{equation}
\sum_{j=0}^{m-1}\sum_{k=0}^{n-1} e^{i\theta_{jk}} |j\rangle_{c1} |k\rangle_{c2} |Q_{jk}\rangle .
\end{equation}
He now measures his part of the state in the $|X_{rs}\rangle$ basis and announces his result, which we will take to be $|X_{rs}\rangle$.  The state is then
\begin{equation}
\sum_{j=0}^{m-1}\sum_{k=0}^{n-1} e^{i\theta_{jk}} e^{-2\pi ijr/m} e^{-2\pi iks/n} |Q_{jk}\rangle  .
\end{equation}
Each party in group 1 now applies the operator $U_{r}$ to their qubit, where $U_{r}|0\rangle = |0\rangle$ and $U_{r}|1\rangle = e^{2\pi ir/m} |1\rangle$, and those in group 2 apply $U_{s}$ to their qubits, where $U_{s}|0\rangle = |0\rangle$ and $U_{s}|1\rangle = e^{2\pi is/n} |1\rangle$.  This will remove the unwanted phase factors in the above sum.  This procedure allows Charlie to implement the operation
\begin{equation}
\sum_{j=0}^{m-1}\sum_{k=0}^{n-1} |Q_{jk}\rangle \rightarrow \sum_{j=0}^{m-1}\sum_{k=0}^{n-1}  e^{\theta_{jk}} |Q_{jk}\rangle .
\end{equation}

Looking at a simple example is useful.  Suppose $m=2$ and $n=3$, which means that the states $|Q_{jk}\rangle$ are three qubit states.  Charlie has been sent the three qubit state $\prod_{l=1}^{3} (\alpha_{l} |0\rangle + \beta_{l}|1\rangle)$. He wants to perform a one-qubit gate on the first qubit and a two-qubit gate, in particular a CPHASE($\theta_{2}$) gate, on the second and third.  He entangles the qubits with his state, and then returns the qubits to the parties that sent them.  At this point we have the states $|Q_{jk}\rangle$
\begin{eqnarray}
|Q_{00}\rangle = \alpha_{1}\alpha_{2}\alpha_{3}|000\rangle & \hspace{5mm} & |Q_{10}\rangle = \beta_{1}\alpha_{2}\alpha_{3} |100\rangle  \nonumber \\
|Q_{02}\rangle = \alpha_{1}\beta_{2}\beta_{3} |011\rangle & \hspace{5mm} & |Q_{12}\rangle = \beta_{1}\beta_{2}\beta_{3} |111\rangle  ,
\end{eqnarray}
and
\begin{eqnarray}
|Q_{01}\rangle & = & \alpha_{1}(\alpha_{2}\beta_{3} |001\rangle +\beta_{2}\alpha_{3}  |010\rangle ) \nonumber \\
|Q_{11}\rangle & = & \beta_{1}( \alpha_{2}\beta_{3} |101\rangle + \beta_{2}\alpha_{3} |110\rangle  ) .
\end{eqnarray}
The first qubit is group 1 and the second two qubits are group 2.  Charlie can use the scheme just discussed to multiply any or all of the qubit states $|Q_{jk}\rangle$ by phase factors.  Charlie now applies the transformation that has the effect
\begin{eqnarray}
|Q_{02}\rangle \rightarrow e^{i\theta_{2}}|Q_{02}\rangle & \hspace{5mm} & |Q_{10}\rangle \rightarrow e^{i\theta_{1}}|Q_{10}\rangle \nonumber \\
|Q_{11}\rangle \rightarrow e^{i\theta_{1}}|Q_{11}\rangle & \hspace{5mm} & |Q_{12}\rangle \rightarrow e^{i(\theta_{1}+\theta_{2}) }|Q_{12}\rangle ,
\end{eqnarray}
and acts as the identity on the remaining $|Q_{jk}\rangle$ states and measures in the $\{ |X_{rs}\rangle \}$ basis.  He broadcasts the result of his measurement, and the three parties perform the appropriate correction operations. The final qubit state is
\begin{eqnarray}
(\alpha_{1} |0\rangle + \beta_{1} e^{i\theta_{1}} |1\rangle )[ \alpha_{2}\alpha_{3}  |00\rangle + \alpha_{2}\beta_{3} |01\rangle \nonumber \\
+ \beta_{2}\alpha_{3} |10\rangle  + e^{i\theta_{2}} \beta_{2}\beta_{3} |11\rangle ] .
\end{eqnarray}
We see that a single qubit operation has been applied to the first qubit, and a CPHASE($\theta_{2}$) to the second and third.

\section{Gate on general qubit states using teleportation}
So far we have looked at the case in which Alice and Bob send Charlie their qubits, Charlie entangles them with his state, and returns them to Alice and Bob.  He then performs operations on his state and communicates classically with Alice and Bob to implement the gate acting on their qubits.  We want to eliminate the step in which Alice and Bob send their qubits to Charlie and replace it with teleportation.  

We start with Alice having the qubit $|\psi_{A}\rangle_{a^{\prime}} = \alpha |0\rangle_{a^{\prime}} + \beta |1\rangle_{a^{\prime}}$, and Bob having the qubit $|\psi_{B}\rangle_{b^{\prime}} = \mu |0\rangle_{b^{\prime}} + \nu |1\rangle_{b^{\prime}}$.  We want to perform a CPHASE($\theta$) gate between them.  Charlie now has a four dimensional system and two qubits, and starts by creating the state 
\begin{eqnarray}
|\Psi\rangle_{abc} & = & \frac{1}{2} ( |0\rangle_{c}|00\rangle_{ab} + |1\rangle_{c}|01\rangle_{ab} \nonumber \\
& & +|2\rangle_{c}|10\rangle_{ab} + |3\rangle_{c}|11\rangle_{ab}) ,
\end{eqnarray}
and sends qubit $a$ to Alice and qubit $b$ to Bob.  Alice wants to transfer the state of her qubit $a^{\prime}$ to an auxiliary qubit $d$, which will be entangled with Charlie, and Bob wants to transfer the state of his qubit $b^{\prime}$ to an auxiliary qubit $e$, which will also be entangled with Charlie.  Let's look at Alice first.  Alice takes an auxiliary qubit $d$ in the state $|0\rangle_{d}$ and sends it and qubit $a$ through a CNOT gate, with qubit $a$ as the control.  Her next step is to measure qubits $a$ and $a^{\prime}$ in the Bell basis
\begin{eqnarray}
|\Psi_{\pm}\rangle_{aa^{\prime}}&  =  &\frac{1}{\sqrt{2}}(00\rangle_{aa^{\prime}} \pm |11\rangle_{aa^{\prime}} ) \nonumber \\
|\Phi_{\pm}\rangle_{aa^{\prime}} & = & \frac{1}{\sqrt{2}}(01\rangle_{aa^{\prime}} \pm |10\rangle_{aa^{\prime}} ) .
\end{eqnarray}
The state before the Bell basis measurement is $1/(2\sqrt{2})$ times the state
\begin{eqnarray}
 [\alpha (|000\rangle_{bcd} + |110\rangle_{bcd})+ \beta (|021\rangle_{bcd} + |131\rangle_{bcd}] |\Psi_{+}\rangle_{aa^{\prime}} \nonumber \\
+ [\alpha (|000\rangle_{bcd} + |110\rangle_{bcd})- \beta( |021\rangle_{bcd} + |131\rangle_{bcd}] |\Psi_{-}\rangle_{aa^{\prime}} \nonumber \\
+[\alpha (|021\rangle_{bcd} + |131\rangle_{bcd})+ \beta( |000\rangle_{bcd} + |110\rangle_{bcd}] |\Phi_{+}\rangle_{aa^{\prime}} \nonumber \\
+[-\alpha (|021\rangle_{bcd} + |131\rangle_{bcd})+ \beta( |000\rangle_{bcd} + |110\rangle_{bcd}] |\Phi_{-}\rangle_{aa^{\prime}} \nonumber
\end{eqnarray}
From this equation we can see what $bcd$ state we will have if we obtain a particular result for the Bell state measurement.  Note that the $bcd$ state in the second line can be transformed into the $bcd$ state in the first line by applying $\sigma_{zd}$, and similarly the $bcd$ state in the fourth line can be transformed into that in the third line also by applying $\sigma_{zd}$.  Finally, the $bcd$ state in the third line can be transformed to that in the first line by first applying $\sigma_{xd}$ and then applying $U_{1c}$ to Charlie's state, where $U_{1c}|2\rangle_{c} = |0\rangle_{c}$, $U_{1c}|0\rangle_{c} = |2\rangle_{c}$, $U_{1c}|1\rangle_{c} = |3\rangle_{c}$, and $U_{1c}|3\rangle_{c} = |1\rangle_{c}$.  Therefore, all of the states resulting from the Bell basis measurement can be transformed by LOCC into the state in the first line, so we can confine the rest of our analysis to that state.

Now Bob does essentially the same thing as Alice.  He takes an auxiliary qubit $e$ in the state $|0\rangle_{e}$ and sends it and the qubit $b$ into a CNOT gate, with $b$ as the control.  He then performs a Bell state measurement on the qubits $b$ and $b^{\prime}$.  The state before the second Bell state measurement is (assuming we got $|\Psi_{+}\rangle_{aa^{\prime}}$ for the first) is $1/(2\sqrt{2})$ times
\begin{eqnarray}
 [\alpha\mu |000\rangle_{cde} + \beta\mu |210\rangle_{cde} + \alpha\nu |101\rangle_{cde} + \beta\nu |311\rangle_{cde} ] |\Psi_{+}\rangle_{bb^{\prime}} \nonumber \\
+ [\alpha\mu |000\rangle_{cde} + \beta\mu |210\rangle_{cde} - \alpha\nu |101\rangle_{cde} - \beta\nu |311\rangle_{cde} ] |\Psi_{-}\rangle_{bb^{\prime}} \nonumber \\
+ [\alpha\mu |101\rangle_{cde} + \beta\mu |311\rangle_{cde} + \alpha\nu |000\rangle_{cde} + \beta\nu |210\rangle_{cde} ] |\Phi_{+}\rangle_{bb^{\prime}} \nonumber \\
+ [-\alpha\mu |101\rangle_{cde} - \beta\mu |311\rangle_{cde} + \alpha\nu |000\rangle_{cde} + \beta\nu |210\rangle_{cde} ] |\Phi_{-}\rangle_{bb^{\prime}} . \nonumber
\end{eqnarray}
Each of the $cde$ states in the last three lines can be transformed into the $cde$ state in the first line by LOCC, so we will assume Bob got $|\Psi_{+}\rangle_{bb^{\prime}}$ for the Bell state measurement.  

In order to implement the CPHASE($\theta$) gate, Charlie can apply the operator $U_{c2}$ to his state, where $U_{c2}$ acts as the identity on $|j\rangle_{c}$ for $j=0,1,2$ and $U_{c2}|3\rangle_{a}=e^{i\theta}|3\rangle_{3}$.  If the phase is supplied to Charlie in the form of a program state (see Eq.\ (\ref{program})), he can transfer it to his state using a Controlled-SHIFT gate as discussed earlier.  He now measures in the basis
\begin{eqnarray}
|v_{1}\rangle_{a} & = & \frac{1}{2} (|0\rangle_{a} + |1\rangle_{a} + |2\rangle_{a} + |3\rangle_{a} )\nonumber \\
|v_{2}\rangle_{a} & = & \frac{1}{2} (|0\rangle_{a} - |1\rangle_{a} - |2\rangle_{a} + |3\rangle_{a} ) \nonumber \\
|v_{3}\rangle_{a} & = & \frac{1}{2} (|0\rangle_{a} + |1\rangle_{a} - |2\rangle_{a} - |3\rangle_{a} ) \nonumber \\
|v_{4}\rangle_{a} & = & \frac{1}{2} (|0\rangle_{a} - |1\rangle_{a} + |2\rangle_{a} - |3\rangle_{a} ).
\end{eqnarray}
If Charlie obtains $|v_{1}\rangle_{a}$, Alice and Bob do nothing, if he gets $|v_{2}\rangle_{a}$, they both apply $\sigma_{z}$, if he gets $|v_{3}\rangle_{a}$, Bob applies $\sigma_{z}$, and if he gets $|v_{4}\rangle_{a}$, Alice applies $\sigma_{z}$. The result is that qubits $d$ and $e$ are in the state resulting from applying the CPHASE($\theta$) gate to two qubits in the state $(\alpha |0\rangle + \beta |1\rangle )(\mu |0\rangle + \nu |1\rangle )$.

\section{Hardware testing}
\label{sec:experiment}

To confirm that the programmable CPHASE($\theta$) protocol can be
realized on current hardware, we implemented it on the IBM Torino
133-qubit superconducting processor. Rather than test the full
teleportation-based version, we implemented the protocol in the
simplified version: Alice and Bob prepare their input qubits
locally, Charlie entangles them with his ancilla, applies the single-qubit
gate $R_z(\theta)$, and then all three qubits are measured. The results of the measurements are used to compute expectation values.  The correction operations are
applied in post processing, in particular, the outcome of
Charlie's measurement is applied as a classical sign correction to
Alice and Bob's data in post-processing. This is the same protocol discussed in the 
paragraph containing Eq.~(8) run start to finish on a single
device, and its success certifies that the measurement mediated
construction works as claimed.

\subsection{Remote protocol implementation}

\begin{figure}[t]
  \centering
  \includegraphics[width=0.9\linewidth]{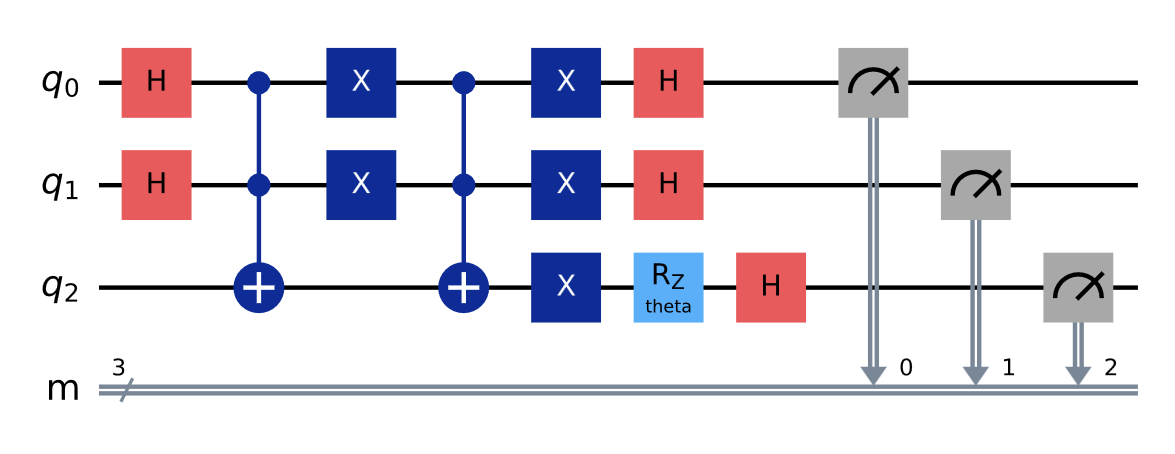}
  \caption{Remote CPHASE($\theta$) protocol implemented on IBM
  Torino. The entangling block between the two CNOTs is
  parameter-independent; only Charlie's $R_z(\theta)$ depends on
  $\theta$.}
  \label{fig:remote}
\end{figure}

The circuit for calculating $\langle X_{A}\rangle$, the expectation value of $\sigma_{X}$
for Alice's qubit, is shown in Fig.~\ref{fig:remote}. The input states for Alice and Bob
(qubits $q_0,q_1$) for the protocol are taken to be $|+x\rangle$ and they are prepared by the two initial Hadamard gates. Charlie's ($q_2$) input is $|0\rangle$. Two
CNOTs, together with the $X$ gates that invert the control state
between them, produce the entangled resource
\begin{equation}
  \lvert\Psi\rangle_{ABC}
  = \tfrac{1}{2}\bigl[\lvert 0\rangle_C(\lvert 00\rangle_{AB}+\lvert 11\rangle_{AB})
  +\lvert 1\rangle_C(\lvert 01\rangle_{AB}+\lvert 10\rangle_{AB})\bigr],
  \label{eq:resource}
\end{equation}
in which Charlie's qubit records the parity of Alice and Bob. The
programmable phase is then imprinted by $R_z(\theta)$ on Charlie's qubit
only, and Charlie then measures his qubit in the $X$ basis (Hadamard followed by
a computational measurement), yielding outcome $c\in\{0,1\}$. Alice
and Bob then measure in the $X$ basis as well, and their raw
expectation values are multiplied by $(-1)^c$ in classical
post-processing. This implements the correction operation.  Because $\theta$ enters only through the single-qubit
rotation on Charlie, the entangling block is independent of $\theta$
and the same resource state works for any value chosen at runtime.

\subsection{Direct baseline}

\begin{figure}[t]
  \centering
  \includegraphics[width=0.65\linewidth]{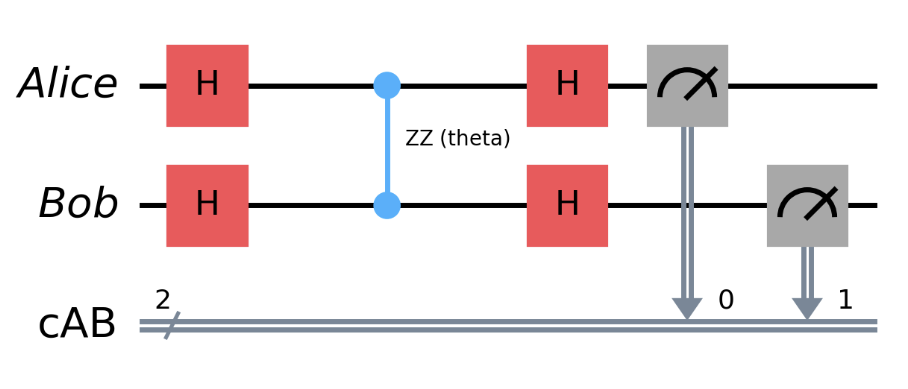}
  \caption{Direct $R_{ZZ}(\theta)$ implementation used as a
  baseline.}
  \label{fig:direct}
\end{figure}

For comparison we also implemented the gate directly with the native
two-qubit rotation $R_{ZZ}(\theta)=\exp(-i\theta Z\otimes Z/2)$
between Alice and Bob (Fig.~\ref{fig:direct}). Both circuits begin
with $|+x\rangle_{a}|+x\rangle_{b}$ and end with $X$-basis measurements, so the
only difference is how the controlled phase is applied. The direct
circuit has $\sim 5$ gate layers against $7$ for the remote protocol,
and serves as a hardware-realistic reference on the same processor.

For both circuits applied to $|+x\rangle_{a}|+x\rangle_{b}$, the ideal prediction
is
\begin{equation}
  \langle X_A\rangle = \cos\theta,
  \label{eq:ideal}
\end{equation}
which provides a one-parameter fringe that cleanly reveals any loss
of coherence.

\subsection{Results}

\begin{figure}[t]
  \centering
  \includegraphics[width=0.9\linewidth]{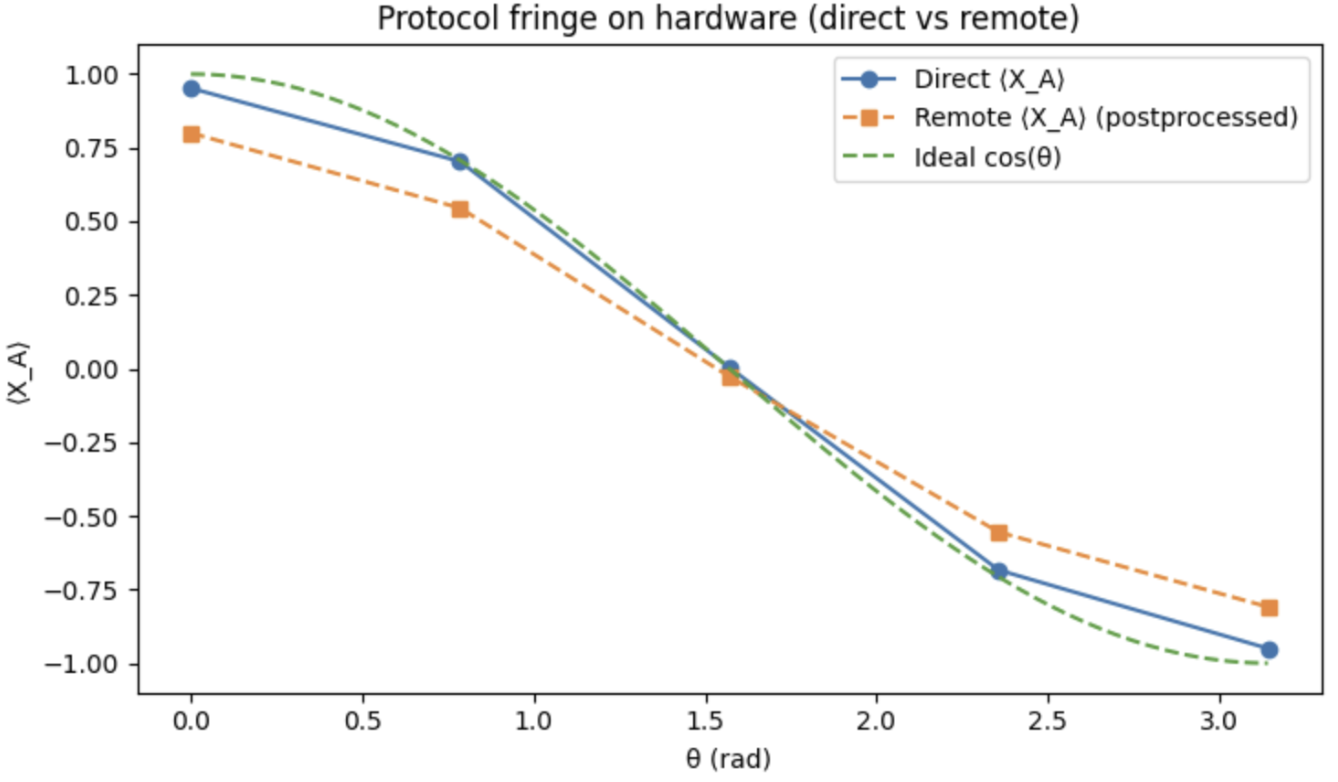}
  \caption{Measured $\langle X_A\rangle$ as a function of $\theta$
  on IBM Torino. Both the direct baseline (blue) and the remote
  protocol with $(-1)^c$ post-processing (orange) follow the ideal
  $\cos\theta$ curve (dashed).}
  \label{fig:fringe}
\end{figure}

Each circuit was run at six values of $\theta\in[0,\pi]$ with
$8000$ shots per point; dynamical decoupling and Pauli twirling were
enabled. Both datasets follow the expected cosine in
Eq.~(\ref{eq:ideal}) (Fig.~\ref{fig:fringe}). Fitting
$\langle X_A\rangle = V\cos(\theta+\phi)+c$ gives
\begin{center}
\begin{tabular}{lccc}
\hline
Protocol & $V$ & $\phi$ (rad) & $c$ \\
\hline
Direct & $0.961\pm0.003$ & $-0.007\pm0.008$ & $\phantom{-}0.002\pm0.003$ \\
Remote & $0.796\pm0.004$ & $\phantom{-}0.018\pm0.010$ & $-0.002\pm0.004$ \\
\hline
\end{tabular}
\end{center}
The phase offsets are consistent with zero ($|\phi|<0.02$ rad in
both cases), confirming that $R_z(\theta)$ correctly encodes the
programmed angle without per-$\theta$ recalibration. As a consistency
check we also measured the two-qubit correlators across all
$\theta$ and found $\langle XX\rangle\approx 1$ and
$\langle YY\rangle\approx 0$, as expected for $R_{ZZ}(\theta)$
acting on $\lvert +x,+x\rangle$.

The ratio $V_r/V_d = 0.83$ isolates the additional coherence cost of
the measurement-mediated implementation: about $17\%$ beyond the
direct gate. This is consistent with the extra circuit depth of the
remote protocol (two additional CNOTs, an extra measurement, and
the associated idle decoherence), and agrees with independent state
tomography at $\theta=\pi/2$, which gave fidelities
$F_d = 0.971\pm0.005$ and $F_r = 0.858\pm0.008$ to the ideal output.

We emphasize that the purpose of this experiment is not to compete
with the native $R_{ZZ}$ gate on fidelity, on a single device the
direct implementation is always cheaper, but to verify that the
programmable protocol produces the correct $\theta$-dependent
two-qubit state when Charlie's measurement is decoded as prescribed.
The agreement with the $\cos\theta$ fringe, the vanishing phase
offset across the full $[0,\pi]$ range, and the tomographic
fidelities together confirm that the measurement-mediated
construction works correctly on real hardware.

To provide a more complete verification of the output state, we
computed additional two-qubit expectation values for both circuits.
The state produced by $R_{ZZ}(\theta)$ acting on
$|+x\rangle_{a}|+x\rangle_{b}$ has the following ideal correlator structure:
\begin{align}
  \langle Z_A Z_B\rangle &= 0, \label{eq:zz}\\
  \langle X_A Y_B\rangle &= 0, \label{eq:xy}\\
  \langle Y_A Z_B\rangle &= \sin\theta. \label{eq:yz}
\end{align}
The first two are zero for all $\theta$, providing a check that
the protocol does not introduce spurious correlations. The third
is a nonzero, $\theta$-dependent two-qubit correlator that tests
whether the programmable phase produces the correct entangled
state---a stronger verification than single-qubit observables alone.

We measured all three correlators at thirteen values of
$\theta\in[0,\pi]$ with $8000$ shots per point for both the direct
and remote circuits. The correlators $\langle Z_A Z_B\rangle$ and
$\langle X_A Y_B\rangle$ are consistent with zero across all
$\theta$, with mean absolute values below $0.015$ for both protocols,
within the expected shot-noise fluctuations of
$1/\sqrt{N_{\text{shots}}}\approx 0.011$.

\begin{figure}[t]
  \centering
  \includegraphics[width=0.9\linewidth]{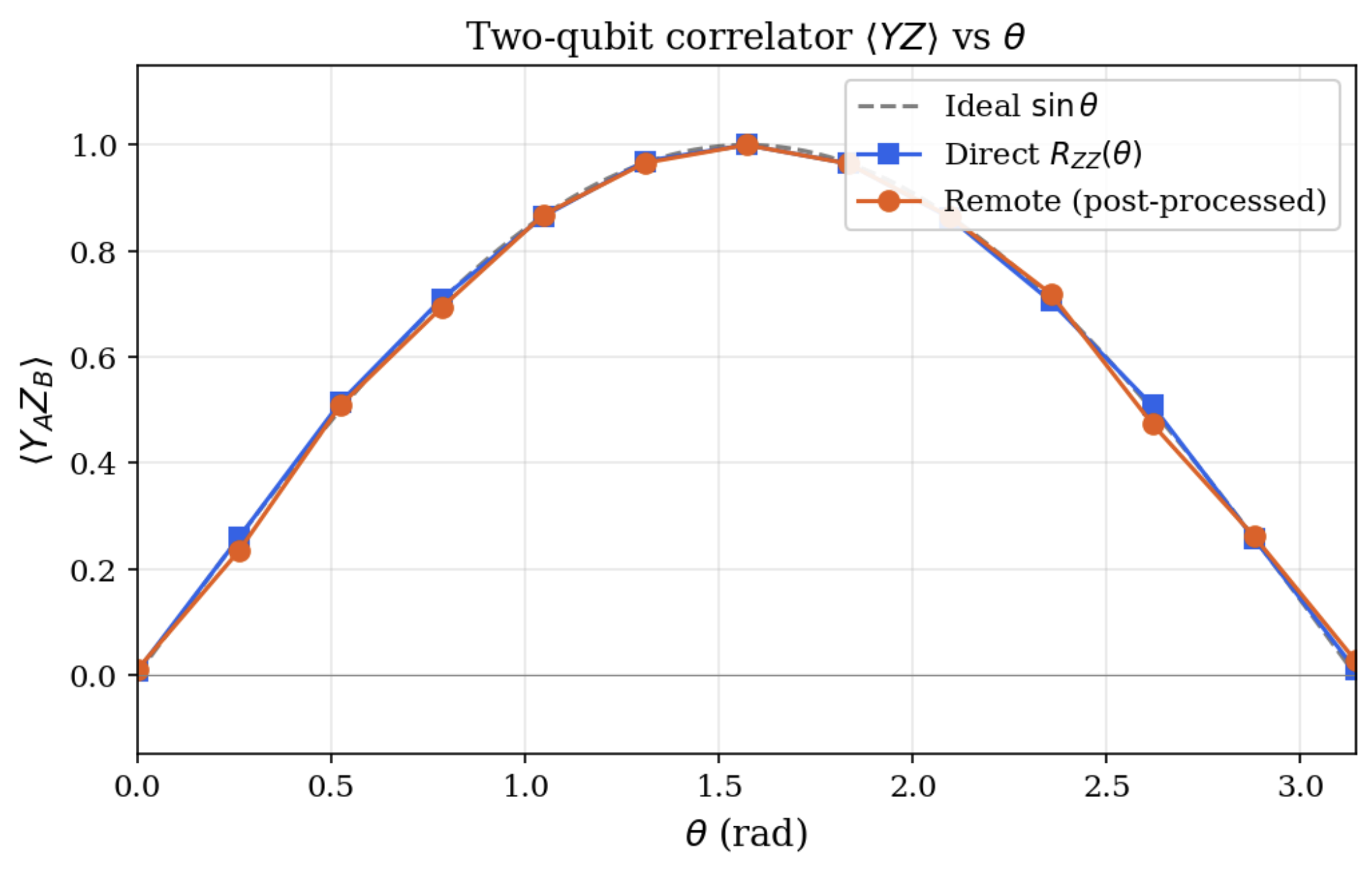}
  \caption{Two-qubit correlator $\langle Y_A Z_B\rangle$ as a
  function of $\theta$. Both the direct baseline (blue squares) and
  the remote protocol with $(-1)^c$ post-processing (orange circles)
  follow the ideal $\sin\theta$ curve (dashed). This confirms that
  the protocol produces the correct $\theta$-dependent two-qubit
  state.}
  \label{fig:yz}
\end{figure}

The two-qubit correlator $\langle Y_A Z_B\rangle$ follows the
predicted $\sin\theta$ dependence for both circuits
(Fig.~\ref{fig:yz}). At $\theta = \pi/2$, both protocols yield
$\langle Y_A Z_B\rangle = 1.000$, matching the ideal value of
$\sin(\pi/2) = 1$. At the endpoints $\theta = 0$ and $\theta = \pi$,
the measured values are within $\pm 0.03$ of zero. Across all
thirteen points, the mean deviation from the ideal $\sin\theta$
curve is $0.004$ for the direct circuit and $0.010$ for the remote
protocol.

Together with the single-qubit fringe $\langle X_A\rangle =
\cos\theta$ (Fig.~\ref{fig:fringe}), the $\langle Y_A Z_B\rangle =
\sin\theta$ correlator provides complementary evidence that the full
$\theta$-dependent state is correctly produced. The former tests a
single-qubit marginal, while the latter probes a genuinely two-qubit
correlation that varies with the programmed angle. The vanishing of
$\langle Z_A Z_B\rangle$ and $\langle X_A Y_B\rangle$ further
confirms that no spurious correlations are introduced by the
measurement-mediated construction.  Summarizing, we see that
the agreement with the $\cos\theta$ fringe, the $\sin\theta$
two-qubit correlator, the vanishing phase offset across the full
$[0,\pi]$ range, the null correlators $\langle ZZ\rangle$ and
$\langle XY\rangle$, and the tomographic fidelities together confirm
that the measurement-mediated construction works correctly on real
hardware.

\section{Conclusion}
The teleportation of quantum gates is a well-established tool of quantum information processing.  Here we have shown that it can be extended to cases in which information about the gate is encoded in the form of a quantum program.   We focussed on the CPHASE($\theta$) gate, but discussed other possibilities as well.  We looked at the case where Alice and Bob, who provide the inputs, sent Charlie, who implements the gate, their qubits, which are then entangled with Charlie's state, and at the case where Alice and Bob use teleportation to entangle their qubits with Charlie's state.  In both cases Charlie is able to implement the CPHASE($\theta$) at a later time either with his choice of $\theta$ or with a value of $\theta$ encoded in a program state.  We also studied multi-qubit gates and discussed under what conditions they can be implemented using only local correction operations.  Finally, we implemented one of our protocols on a quantum computer and showed that it was possible to run it on existing hardware.

\section*{Acknowledgment}
This research was supported by the National Science Foundation under the grant Collaborative Research: NeTS: Medium 2504622.

\end{document}